\documentstyle[agupp]{article}

\input{epsf}

\lefthead{PALMER ET AL.}
\righthead{AN INVERSE METHOD FOR RADIO OCCULTATION DATA}
\received{August, 1999}
\revised{January, 2000}
\accepted{February, 2000}
\paperid{JGRd-1999R462}
\cpright{AGU}{1999}
\ccc{0148-0227/96/96JA-00000\$09.00}

\authoraddr{Dr. Paul I. Palmer,
Division of Engineering \& Applied Science,
Pierce Hall, 
Harvard University, 
Cambridge, MA 02138
(e-mail: pip@io.harvard.edu)
}

\slugcomment{To appear in the {\it Journal 
of Geophysical Research}, 2000.}

\begin{document}

\title{A non-linear optimal estimation inverse method for radio occultation 
       measurements of temperature, humidity and surface pressure}

\author{Paul I. Palmer\altaffilmark{1},
        J. J. Barnett}

\affil{Department of Physics, 
       Clarendon Laboratory, 
       Oxford, 
       United Kingdom}

\author{J. R. Eyre, S. B. Healy}

\affil{Satellite Applications Division,
       United Kingdom Meteorological Office,
       Bracknell,
       United Kingdom}

\altaffiltext{1}{Now at Division of Engineering \& Applied Science, 
                 Pierce Hall,  
                 Harvard University,
                 Cambridge, MA 02138}

\begin{abstract}
An optimal estimation inverse method is presented which can be used to retrieve
simultaneously vertical profiles of temperature and specific humidity, in 
addition to surface pressure, from satellite-to-satellite radio occultation 
observations of the Earth's atmosphere.
The method is a non-linear, maximum {\it a posteriori} technique which can 
accommodate most aspects of the real radio occultation problem and is found 
to be stable and to converge rapidly in most cases. The optimal estimation 
inverse method has two distinct advantages over the analytic inverse 
method in that it accounts for some of the effects of horizontal gradients and 
is able to retrieve optimally temperature and humidity simultaneously from the 
observations. It is also able to account for observation noise and other 
sources of error. Combined, these advantages ensure a realistic retrieval of 
atmospheric quantities.

A complete error analysis emerges naturally from the optimal estimation theory,
allowing a full characterisation of the solution. Using this analysis a 
quality control scheme is implemented which allows anomalous retrieval 
conditions to be recognised and removed, thus preventing gross retrieval 
errors.

The inverse method presented in this paper has been implemented for bending 
angle measurements derived from GPS/MET radio occultation observations of the 
Earth. Preliminary results from simulated data suggest that these observations 
have the potential to improve NWP model analyses significantly throughout 
their vertical range. 
\end{abstract}


\begin{article}

\section{Introduction}

Radio occultation (RO) experiments have played a prominent role in the NASA 
programme for solar system exploration for more than two decades and have 
contributed to studies of the atmosphere of 
Mars [\markcite{{\it Fjeldbo and Eshleman,} 1968}], 
Venus [\markcite{{\it Fjeldbo and Kliore,} 1971}],
Jupiter [\markcite{{\it Kliore et al,} 1975}],
Saturn [\markcite{{\it Lindal et al,} 1985}],
Uranus [\markcite{{\it Lindal et al,} 1987}],
and Neptune [\markcite{{\it Lindal} 1992}]. This method of radio occultation
uses a receiver on Earth and a satellite occulted by a planetary atmosphere
(which may occur from either a fly-by or by a satellite orbit of the planet).
Suitably accurate atmospheric RO measurements of the Earth's atmosphere
became possible with the advent of the Global Positioning System (GPS),
but it was not until the late 1980s$-$early 1990s that the potential of RO 
using the GPS was widely appreciated 
(e.g. \markcite{{\it Gurvich and Krasil'nikova,}} [1990]).

The radio occultation method used to sound the Earth's atmosphere is different 
from that used by the planetary experiments, in that both the receiver and the
transmitters are orbiting the planet.

Data from the prototype GPS space-borne receiver, GPS/MET, launched in April 
1995, confirmed the potential of obtaining accurate, global observations of 
the Earth's atmosphere from the radio occultation technique. Temperature 
comparisons between early results from the GPS/MET receiver and collocated 
radiosondes and numerical weather prediction (NWP) model analyses showed good 
agreement (e.g. \markcite{{\it Kursinski et al,}} [1996]; 
\markcite{{\it Ware et al,}} [1996]).

The analytic method of inverting radio occultation measurements to 
obtain meteorological parameters (i.e.~the method used to sound other planetary
atmospheres) involves the use of an integral transform, using the assumption
of a horizontally homogeneous atmosphere, to obtain a profile of refractivity
(as a function of geometric height) 
[\markcite{{\it Fjeldbo and Eshleman,} 1968}]. 
The hydrostatic relation is used to obtain pressure and temperature from 
refractivity via density.
For the Earth's atmosphere, where a reasonable prior knowledge of horizontal
gradients is available, the analytic inversion does not represent the most 
suitable method since inadequate modelling of such gradients can cause large 
retrieval errors (e.g. \markcite{{\it Ahmad and Tyler,}} [1998]).

\markcite{{\it Eyre,}} [1994] addresses this issue and suggests a 
statistically optimal retrieval approach, using variational methods, to 
enable the direct assimilation of bending angle or refractivity 
(\markcite{{\it Healy and Eyre,}} [1999] investigate the latter quantity). 
\markcite{{\it Zou et al,}} [1995] also looked at the impact of atmospheric
radio refractivity measurements using a 4-D variational data assimilation
approach. Their results showed that the measurements were effective in 
recovering the vertical profiles of water vapour, and found that the accuracy
of the derived water vapour field was significantly better than that obtained
through the analytic retrieval technique. The assimilation of these 
measurements were also shown to provide useful temperature information.
There have also been several numerical experiments which have assessed 
simulated GPS/MET refractivity measurements to predict cyclonic disturbances
(e.g. \markcite{{\it Kuo et al,}} [1998]), and have concluded that these
measurements are likely to have a significant impact on short-range operational
NWP, with the caveat that the number of GPS receivers will have to be 
increased before the full potential impact of this measurement could be 
realised.

In this paper we utilise a non-linear optimal estimation technique which is 
implemented and validated using an ensemble of simulated retrieval scenarios,
using the bending angle quantity as the `observation'.

Section \ref{sect:RO-measurements} outlines the details of the RO methodology 
for the Earth's atmosphere necessary to derive the bending angle quantity, 
recalls the analytic inverse method and discusses the impetus for pursuing 
an alternative inverse method. In section \ref{sect:optimal} we 
outline the theory for the non-linear optimal estimation inverse method and 
give details of its implementation for GPS RO observations. Section 
\ref{sect:results} is devoted to details of the validation of the optimal 
estimation inverse method with reference to GPS RO observations. The 
sensitivity of the inverse model assumptions is also investigated, and we 
conclude the paper with a discussion of the results obtained.

\section{Radio occultation measurements of Earth's atmosphere}
\label{sect:RO-measurements}

\markcite{{\it Kursinski et al,}} [1997] give a detailed description of the 
method used to measure the RO atmospheric observables; the following section 
gives a summary of the theory, assuming no external encryption of the signals.

The GPS satellites transmit on two L-band radio frequencies\footnote{Namely, 
L1: 1.57542~GHz and L2: 1.2272~GHz.}. Assuming a continuous link between the 
receiver and transmitter, when the receiver passes behind the atmosphere with 
respect to a GPS transmitter the signal travels through the 
atmosphere and is refracted in response to variations of refractive index 
along its path. This refraction causes the ray to travel over a longer path 
than it would in the absence of the atmosphere, in accordance with 
Fermat's principle of least time, which subsequently causes an atmospheric 
time delay in the received signal. 

\begin{figure}
\centerline{
\epsfxsize=16cm
\epsffile{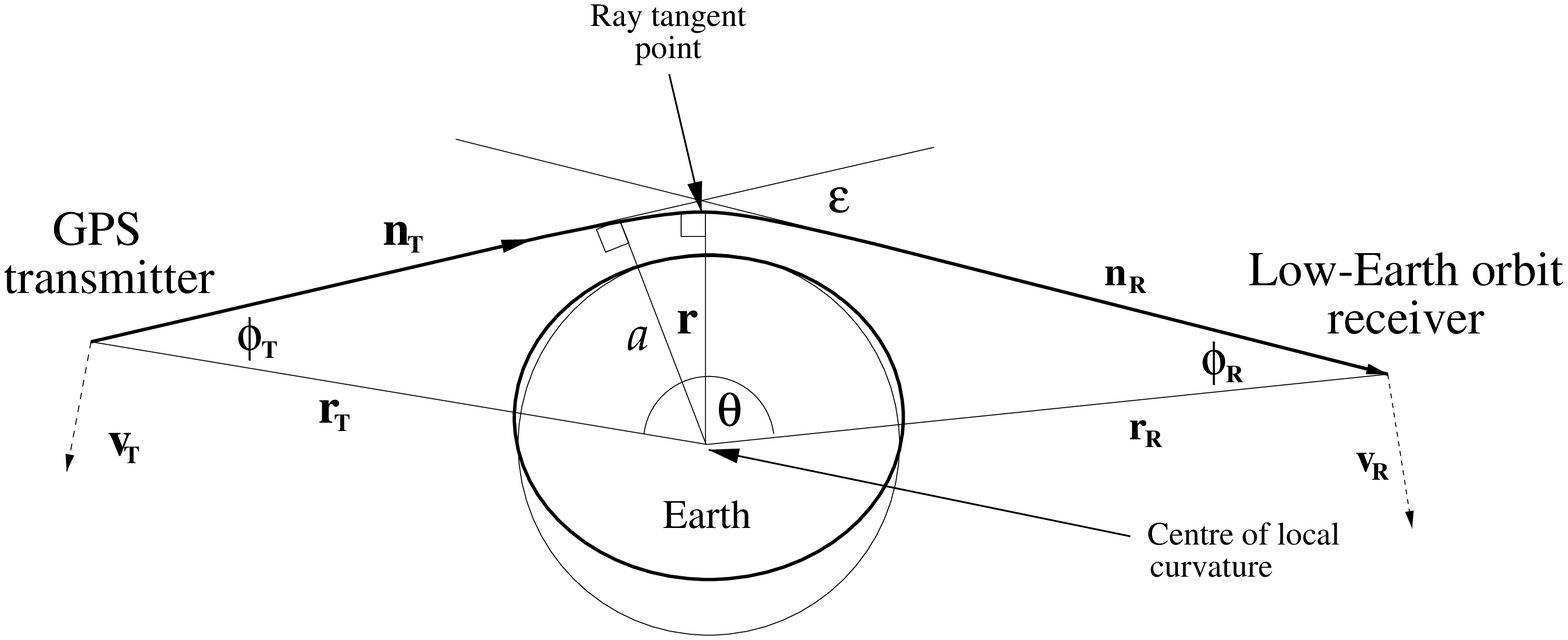}}
 \caption{Defining the radio occultation geometry used to obtain 
          bending angle information from the time delay caused
          by the Earth's atmosphere. A tangential sphere is superimposed on to 
          the oblate Earth (exaggerated) to emphasise the position of the
          local radius of curvature at the ray periapsis. The dashed lines 
          indicate motion.}
\label{fig:time-eps}
\end{figure}

The Doppler shift of the signal is calculated from the additional atmospheric 
delay (the derivative of the phase delay). Using the geometry and notation of 
figure \ref{fig:time-eps}, the Doppler shift $f_{\rm d}$ of the carrier 
frequency $f_{0}$ measured by the receiver is given by:

\begin{equation}
f_{\rm d} =  f_{0} \left[
                    \frac{({\bf v}_{\rm T}\cdot{\bf n}_{\rm T} + 
                    {\bf v}_{\rm R}\cdot{\bf n}_{\rm R})}{c} \right]
\label{eq:fd}
\end{equation}
where ${\bf v}_{\rm T,R}$ are transmitter and receiver velocity vectors, 
${\bf n}_{\rm T,R}$ are path direction vectors of the transmitter and receiver,
and $c$ is the velocity of light. 

There are also relativistic terms which need to be considered in equation 
\ref{eq:fd} (due to different gravitational potentials and higher order 
corrections for spacecraft velocity) but these can be eliminated based
on knowledge of orbital geometry and the Earth's gravity field 
[\markcite{{\it Kursinski et al,} 1997}]. Note that the relative positions 
and velocities of the two satellites can be calculated very accurately using 
available tracking data, which is independent of radio occultation data.

By specifying radial and tangential components of the velocity of satellite 
$i$ in the plane coinciding with the ray trajectory as ${\bf v}^{\rm r}_{i}$ 
and ${\bf v}^{\rm t}_{i}$, and taking into account Snell's law, the angles 
$\phi_{\rm R}$ and $\phi_{\rm T}$ can be calculated from the following 
relations:

\vspace{-0.5cm}
\begin{eqnarray}
f_{\rm d} &=& f_{0}c^{-1}({{\bf v}^{\rm r}_{\rm R}}\cos\phi_{\rm R} + 
              {\bf v}^{\rm r}_{\rm T}\cos\phi_{\rm T} + \nonumber \\
          & & {\bf v}^{\rm t}_{\rm R}\sin\phi_{\rm R} -
              {\bf v}^{\rm t}_{\rm T}\sin\phi_{\rm T})
\label{eq:comp_delay}
\end{eqnarray}

The cumulative effect of the atmosphere on the ray path can be expressed in 
terms of the total refractive bending angle, $\varepsilon$, as a function of 
the impact parameter, ${a}$. The impact parameter may be defined as the 
perpendicular distance between the local curvature of the Earth at the 
tangent point of the ray and the asymptotic straight line followed by the ray 
as it approaches the atmosphere.

From Bouguer's rule [\markcite{{\it Born and Wolf,} 1993}], and the geometry 
defined by figure \ref{fig:time-eps}, $\varepsilon(a)$ can be calculated thus 

\vspace{-0.5cm}
\begin{eqnarray}
r_{\rm R}\sin\phi_{\rm R} = r_{\rm T}\sin\phi_{\rm T} = a \\
\varepsilon = \phi_{\rm R} + \phi_{\rm T} + \theta - \pi 
\label{eq:bendingangle}
\end{eqnarray}
It is this rule that introduces the assumption of spherical symmetry
($nr\sin\phi = a$, where $a$ is a constant along the ray path), i.e. 
a horizontally homogeneous atmosphere. Departures from this assumption
can introduce significant errors if not properly accounted for. These errors 
have been studied by \markcite{{\it Ahmad and Tyler,}} [1999] and
\markcite{{\it Healy,}} [1999], but are not addressed in the study presented 
here.

Measurements of the time delay become possible for neutral atmospheric 
sounding when the GPS signal begins to transect the mesosphere at an altitude 
of about 85~km; at this altitude the atmospheric phase delay is about
1~mm (3 $\times$ 10$^{-12}$~s) which can be observed by the LEO GPS receiver 
[\markcite{{\it Ware et al,} 1996}]. Further information about the 
measurement characteristics may be obtained from 
\markcite{{\it Kursinski et al,}} [1997].

\subsection{The analytic inverse method}

Using an Abel integral transform (equation \ref{eq:Abel}), 
or making use of a similar integral transform when applying Fresnel 
diffraction theory \markcite[{{\it Mortensen and H{\o}eg}, 1998}], these 
bending angle measurements can be inverted to obtain a profile of 
refractivity. For completeness sake, the `forward' Abel integral transform
(equation \ref{eq:Abel2}) is presented alongside the `inverse' Abel integral 
transform:

\begin{equation}
\ln n(x)       = \frac{1}{\pi} \int_{x}^{+{\infty}}
		    {\varepsilon}(a)({a^2 - x^2})^{-1/2}{\rm d}a
\label{eq:Abel}
\end{equation}

\begin{equation}
\varepsilon(a) = -2a \int_{a}^{+{\infty}} 
		    \left( \frac{\partial\ln n(x)}{\partial x}\right)
		    ({x^2 - a^2})^{-1/2}{\rm d}x, 
\label{eq:Abel2}
\end{equation}
where $n$ is the refractive index and $x$ is the refractive radius (i.e. 
$x=rn$).

Geometric height levels, $z$, can be obtained from the refractive index
profile, as a function of impact parameter, and the local radius of curvature,
$R_{c}$, thus

\begin{equation}
z = \frac{x}{n} -  R_{c}.
\end{equation}

Because refractive index is near unity, refractivity $N$ is used to 
describe the refractive medium which is given by $N = (n-1)\times10^{6}$.

Refractivity is affected primarily by air density (dependent on pressure 
and temperature) and water vapour density, thus the measurement contains 
information about both.  Equation \ref{eq:refractivity} describes this 
relationship, 

\begin{equation}
N = \frac{c_{1}P_{a}}{T} + \frac{c_{2}P_{W}}{T^{2}}
\label{eq:refractivity}
\end{equation}
where $P_{a}$ is the total atmospheric pressure (hPa), $P_{W}$ is the partial 
pressure of water vapour, $T$ is the temperature (K), and $c_{1}$ and $c_{2}$ 
represent constants of proportionality, whose values are  $77.6$ (KhPa$^{-1}$) 
and $3.73\times10^{5}$ (K$^{2}$hPa$^{-1}$), respectively. The form of the dry
and moist terms in equation \ref{eq:refractivity} is from 
\markcite{{\it Smith and Weintraub,} [1953]}.

Refractivity is also affected by charged particles in the ionosphere and 
the scattering by water droplets suspended in the atmosphere. The first-order
ionospheric contribution to refractivity can be removed by combining the 
two GPS signals (described by \markcite{{\it Vorob'ev and Krasil'nikova,}} 
[1997]), leaving higher-order terms, and the scattering 
contribution is found to be negligible compared to the contribution due to
air and water vapour density \markcite[{{\it Kursinski et al,} 1997}].

There is no measurement information to allow the separation of the effects of
temperature and water vapour, and therefore these quantities can be retrieved 
only using prior information. If the 
absolute humidity is judged to be small (e.g.~in the coldest regions of the
troposphere and stratosphere, with temperatures less than 250~K), 
it may be neglected, and density calculated from 
refractivity. The hydrostatic relation can be used to calculate values of 
pressure, and hence temperature. However, if humidity is judged to be 
significant, then an iterative process may be used to calculate 
temperature/humidity if an {\it a~priori} profile of humidity/temperature is 
used. This prior information can be taken from various sources such as 
collocated NWP model output.

The inability of this inverse method to account for horizontal refractivity
inhomogeneities, and the sub-optimal way this method retrieves temperature
and humidity with prior values, represents two key disadvantages 
of this method, and form part of the impetus to develop a new inverse method.

The hydrostatic relation, used to compute values of pressure on the retrieved
height levels, requires an assumed pressure value at a 
particular geometric height level. 
A variety of methods have been implemented to tackle this initial value 
problem. {\it Kursinski et al,} [1996] assumed a temperature of 260~K at
50~km. This method has the problem that if the assumed temperature is 
inconsistent with the measurements an error is introduced, which decreases
exponentially with depth. More elaborate methods 
(e.g. {\it Rocken et al,} [1997] and {\it Steiner et al,} [1999]) initialise 
the GPS/MET retrieval at some high altitude (e.g. 100~km) using climate model 
data, and combine the measurements and model data to minimise downward 
propagation of errors. In principle, the method presented in this paper
also combines the model data and the observations but achieves this is an
optimal way. The method presented also has the advantage of a straightforward
error characterisation.

Since GPS RO observations sometimes reach near-surface altitudes (i.e. less 
than 1~km from the surface), surface pressure is also retrieved using the 
optimal estimation inverse method.

\section{The optimal estimation inverse method}
\label{sect:optimal}

The method outlined here is also known as one-dimensional variational data 
analysis. 

The main advantage of this method is that it provides simultaneous estimates
of temperature and humidity profiles that are statistically optimal, given
prior estimates from an NWP model (together with their error covariances).
It also provides a framework for assessing the error characteristics of the
estimates.

In this study only the 1-d problem has been studied. However, the errors 
introduced by the neglect of the horizontal gradients have been estimated
and allowed for as part of the error budget (see section 
\ref{ssect:implementation} ``Forward modelling errors'').

\subsection{Theory}

A brief description of the theory used in optimal estimation in presented here;
a more detailed description may be found in \markcite{{\it Rodgers,}} [1976] 
and \markcite{{\it Rodgers,}} [1990].

For brevity, the observation noise, 
the error associated with any forward modelling parameters and the 
forward model error (which includes the representativeness error 
[\markcite{{\it Lorenc,} 1986}] ) will be accounted 
for in one vector which will be denoted by ${\mbox{\boldmath{$\epsilon$}}}$ 
and its ensemble characteristics described by the covariance 
matrix ${\bf E}$. 

The rationale behind optimal estimation is to minimise a cost functional 
${\bf J}({\bf x})$ (or to solve $\nabla_{\rm x}{\bf J}({\bf x})$=0), 
which measures the degree of fit of estimates of the atmospheric state to the 
measurements and to some prior information, and possibly to some other 
physical or dynamical constraints. In this case ${\bf J}({\bf x})$ is given by

\begin{eqnarray}
\label{eq:pntly-fn}
 {\bf J}({\bf x}) &=& 
                   ({\bf y}^{\rm o} - {\bf y(\bf x)})^{\rm T}
                    {\bf E}^{-1}
                   ({\bf y}^{\rm o} - {\bf y(\bf x)}) + \nonumber\\
                  & & ({\bf x} - {\bf x}^{\rm b})^{\rm T}
                    {\bf C}^{-1} 
                   ({\bf x} - {\bf x}^{\rm b})
\end{eqnarray} 
where ${\bf x}^{\rm b}$ and ${\bf x}$ represent the background and updated
state vectors, respectively; ${\bf y}^{\rm o}$ and ${\bf y(\bf x)}$ represent 
the observation vector and the estimated observation vector calculated from
the state vector (\markcite{{\it Eyre,}} [1994]), respectively; and 
${\bf C}$ represents the background error covariance matrix. 

There are a number of methods available to minimise ${\bf J}({\bf x})$: the 
scheme described here uses the Leven\-berg-Marquardt iterative method 
[e.g. \markcite{{\it Press et al,} 1992}]:

\begin{eqnarray}
{\bf x}_{\rm i+1} &=& {\bf x}^{\rm b} +
                      ((1 + \gamma){\bf C}^{-1} + 
                      {\bf K}^{\rm T}{\bf E}^{-1}{\bf K})^{-1} \nonumber \\
                  & & [({\bf K}^{\rm T}{\bf E}^{-1}({\bf y}^{\rm o} - 
                      {\bf y(x_{\rm i})})) + \nonumber \\
                  & &    (\gamma{\bf C}^{-1} + 
          {\bf K}^{\rm T}{\bf E}^{-1}{\bf K})({\bf x}_{\rm i}-{\bf x}^{\rm b})]
\label{eq:marqleven}
\end{eqnarray}
where ${\bf K}$ is $\nabla_{{\rm\bf x}_{\rm i}}{\bf y(x_{\rm i})}$, 
$\gamma$ is a non-dimensional weighting factor
\footnote{for increasing values of $\gamma$ this minimisation method 
degenerates into the method of steepest descent.}, and all other variables 
are as before.

Using the optimal estimation theory it is possible to obtain an 
error covariance for the retrieved products. Indeed, it can be argued that 
the retrieved products are of limited value without an estimate of their 
uncertainty. 
The solution error covariance ${\bf\hat{S}}$ is given approximately (i.e.
at the linear limit) by

\begin{equation}
{\bf\hat{S}} = ({\bf C}^{-1} + 
                {\bf K}^{\rm T}{\bf E}^{-1}~{\bf K})^{-1}.
\label{eq:solutionS}
\end{equation}

The solution error covariance can then be compared with the prior error
covariance to ascertain how the retrieval has improved upon the prior
knowledge of the atmospheric state.

\subsection{Implementation of the optimal estimation inverse model}
\label{ssect:implementation}

This subsection describes in detail the components of equation 
\ref{eq:pntly-fn}.

\subsubsection*{The background state vector and its uncertainty 
                covariance matrix}

In this case the background knowledge of the atmosphere state 
${\bf x}^{\rm b}$ was obtained from short-range forecasts provided by the 
UKMO unified model [\markcite{{\it Cullen,} 1993}]. The model from which the 
data are derived had 19 levels, which were expressed on hybrid-sigma pressure 
coordinates (surface$-$10~hPa). The 
global model had a resolution of 0.833$^{\circ}$ (180$^{\circ}$/217) latitude 
and 1.25$^{\circ}$ (360$^{\circ}$/288) longitude. 

The data used are 6-hour forecasts which have been interpolated to 
occultation event positions, using the mean latitude and longitude of each 
occultation\footnote{The mean latitude and longitude of an occultation 
corresponds typically to altitudes in the lower stratosphere/upper 
troposphere.}.

These 19 levels are linearly interpolated (in $\ln$-pressure) onto the state 
vector levels used for TOVS retrievals [\markcite{{\it Eyre,} 1990}]. 
CIRA climatology [\markcite{{\it CIRA,} 1986}] is assumed
above the UKMO model accounting for the latitudinal and seasonal variation of 
the profile. This climatology provides a reasonable prior and first guess 
information in the upper stratosphere.

For the forecast error covariance matrix ${\bf C}$ 
(described in \markcite{{\it Eyre,}} [1989]), lower atmospheric values
(surface$-$50~hPa) were generated from radiosonde$-$forecast difference
statistics and upper stratosphere values were found by regression from the 
levels provided [\markcite{{\it Eyre,} 1989}]. 

The radio occultation retrieval uses 40 temperature elements, 15 
$\ln$(specific humidity) elements\footnote{Specific humidity is expressed as 
the natural logarithm of specific humidity since forecast errors in this 
quantity are more constant than in specific humidity.} and a surface 
pressure element from this forecast error covariance matrix.
The temperature and $\ln$(specific humidity) inter-quantity covariance values 
have been set to zero. These inter-quantity covariances are not well known and 
assuming zero covariance between them is more conservative than an erroneous 
covariance. The surface pressure element is uncorrelated with both temperature 
and $\ln$(specific humidity).

Since CIRA climatology is used to form the {\it a~priori} (and the first-guess)
it is necessary to 
consider the errors that may be attached to such information. In general, if 
the standard deviation values from the diagonal elements of the forecast error covariance 
matrix are smaller than the uncertainties assumed for the climatology, then 
the climatological errors are used at the levels in the upper atmosphere 
described by the CIRA climatology (off-diagonal elements remain the same): 
at latitude $\theta$, for $|\theta|\geq 45^{\circ}$ $\sigma$=15~K (winter) and 
$\sigma$=5~K (summer); and for $|\theta|\leq 45^{\circ}$ $\sigma$=5~K. 

\begin{figure}[tbh!]
  \epsfysize=7cm
  \centerline{\epsffile{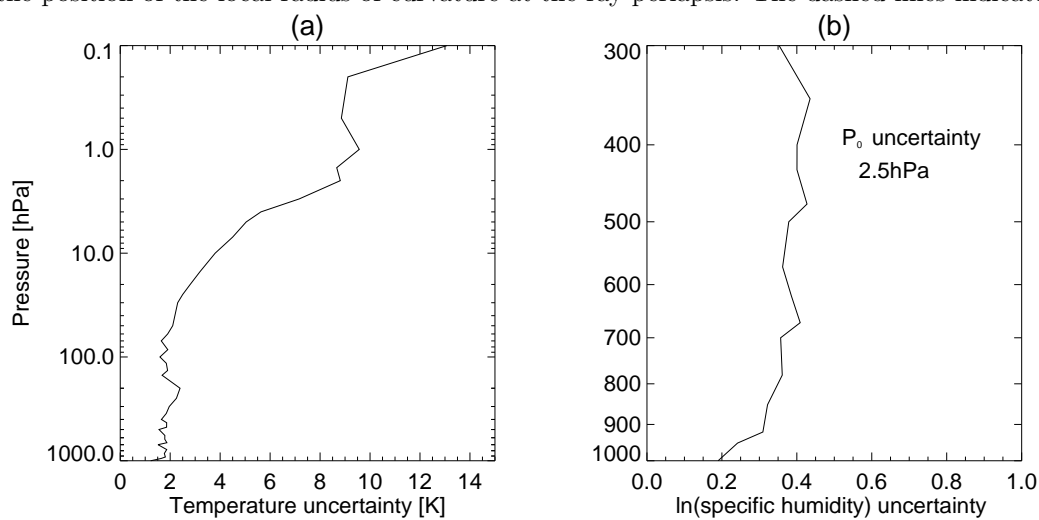}}
  \caption{The square-root values of the principal diagonal of the 
           UKMO forecast error covariance matrix. 
           Panels (a) and (b) show the assumed uncertainty for prior
           temperature and $\ln$(specific humidity), respectively, with the 
           assumed prior uncertainty for the 
           surface pressure element inset of panel (b).}
  \label{fig:ukmoerror}
\end{figure}

The values for the diagonal elements of the UKMO forecast error covariance 
matrix are shown by figure \ref{fig:ukmoerror}.

As expected for temperature, the lower atmosphere forecast errors are 
reasonably small (of the order of 1.5 K) and increase as a function of 
altitude. 
The actual values have been developed over recent years at the UKMO and 
reflect the average error in six-hour forecasts.

\subsubsection*{The observation vector and its error covariance matrix}

The observation vector ${\bf y}^{\rm o}$ contains bending angle measurements as
a function of impact parameter (section \ref{sect:RO-measurements}). 

In practice, atmospheric phase delay measurements from the GPS/MET receiver 
are low-pass filtered to reduce noise. The cut-off frequency of the filter is 
tuned to pass phase variations corresponding to vertical scales of 2 to 3~km 
in the stratosphere and approximately 200~m in the lower troposphere 
[\markcite{{\it Rocken et al,} 1997}]. From the phase observable, Doppler 
shifts (and subsequently bending angle profiles) for the two GPS signals are 
computed. 
First-order ionospheric effects are removed from the data by combining the two
signals to form a single corrected profile [\markcite{{\it Vorob'ev and 
Krasil'nikova,} 1994}]. Typically, after filtering, there are  100$-$200 
neutral atmosphere bending angle measurements, which span a vertical range of 
approximately 0.5$-$60~km.
 
In addition to the observation error covariance matrix consisting of 
observation noise estimates, errors from the forward modelling and forward 
model parameters are considered.

\subsubsection*{Observation noise}

Observation errors are created by the hardware of the measurement system and 
by the pre-processing of the observations. The observation noise estimates are 
taken from the work described in \markcite{{Luntama,}} [1997]. They include 
thermal noise, residual errors from the ionospheric correction, local
multipath (distortions when the transmitted signal is reflected from a surface 
near the signal propagation path), 
orbit determination accuracy, and clock instabilities of low-Earth-orbit 
receiver and GPS satellites and ground stations.

These observation error estimates are based on phase noise levels during the 
measurement or estimated from other noise sources during a radio occultation 
event. The effect from satellite clock errors and from the selective 
availability military encryption process were found to be negligible 
by assuming a differencing decryption technique (involving the differencing 
of the several sets of signals, using the satellites and the ground 
stations) [\markcite{{\it Kursinski et al,} 1997}].
The only exception is the residual error from the ionospheric correction
which was obtained from refractivity error estimates published in 
\markcite{{\it Kursinski et al,}} [1997] and mapped to bending angle space 
using a forward model [\markcite{{\it Luntama,} 1997}]. 

\begin{figure}[tbh!]
  \epsfxsize=14cm
  \centerline{\epsffile{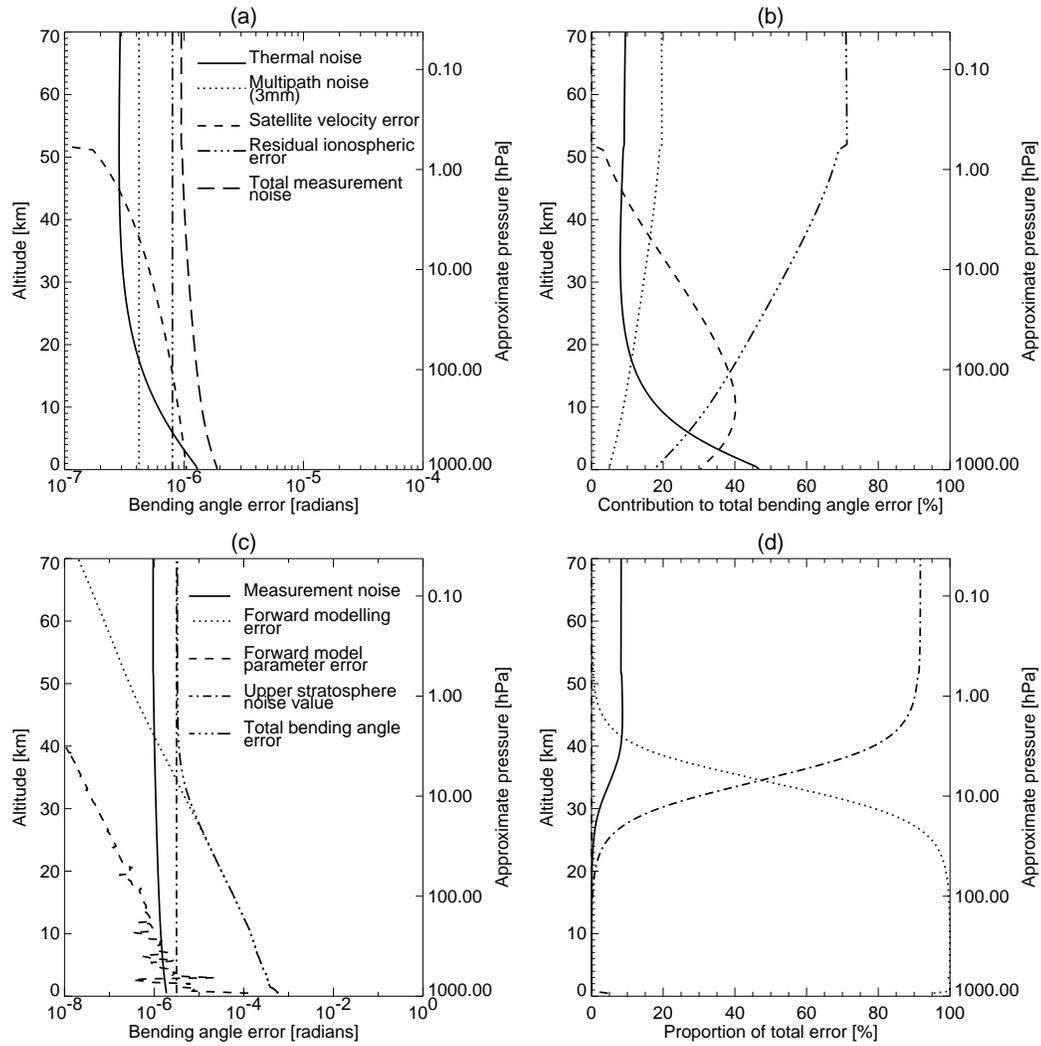}}
  \caption{Measurement noise budget and total measurement error budget for 
           the RO optimal estimation inverse method 
           [after \markcite{{Luntama,}} [1997]]. 
           The error estimates shown in panel (a) represent normal 
           atmospheric conditions, i.e.~small multipath error (3~mm) and 
           normal ionospheric conditions. Panel (b) shows the 
           percentage contributions from the different error sources to the 
           total bending angle error.
           Panel (c) shows typical standard deviation values from the 
           principal 
           diagonal from each contribution to the total measurement error 
           covariance matrix; and panel (d) shows the individual measurement
           error sources as a percentage proportion of the total error.}
  \label{fig:bendingnoise}
\end{figure}

The observation error estimates\footnote{These estimates have been computed 
by a nominal bending angle profile defined using an exponential curve with a
scale height of approximately 7~km [J. P. Luntama: personal 
communication].} used in the optimal estimation technique are 
shown by figure \ref{fig:bendingnoise}. These estimates represent normal 
atmospheric conditions with a relatively small multipath error (3~mm) and 
normal ionospheric conditions. Panel (a) shows 
that there are a number of noise contributions of comparable size in the
lower atmosphere. At altitudes above 30~km the residual ionospheric correction
error begins to dominate the total bending angle error curve (panel (b)).

These noise estimates are assumed to be fully independent, i.e.~their 
inter-level (and inter-quantity) covariance is zero. However, the bending 
angle measurements do contain a small, local correlation between successive 
levels due to filtering of the phase measurements. Because this correlation is 
small, the diagonal form is a good approximation to the full matrix 
[J. P. Luntama: personal communication]. 

Real observations from the GPS/MET data used have been found to be 
noisier than theoretical estimates [\markcite{{\it Luntama,} 1997}]. 
Bending angle fluctuations in the upper stratosphere (of the order of 
$10^{-5}$ radians) are present and are thought to be due to residual errors
from the LEO satellite clock calibration in the differencing decryption 
technique (see ``Observation noise'') [\markcite{{\it Syndergaard,} 1999}]. 
As such, a suitable error is attached to reflect the upper atmosphere 
measurements. 

\subsubsection*{Forward modelling errors}

In this work the forward model used to map from state space to observation 
space is 
described in \markcite{{\it Eyre,}} [1994] but applied to an atmosphere
approximated as spherical symmetric about the given profile at the tangent
point. Essentially the geophysical parameters are 
converted to refractivity as a function of height, and subsequently impact 
parameter using the local radius of curvature. The resulting profile is mapped 
into observation space using the `forward' Abel integral transform 
(equation \ref{eq:Abel2}).

The two main forward modelling errors are due to the assumption of a 
horizontally homogeneous atmosphere and a representativeness error.

Estimates for the first of these errors are obtained using a version 
of the forward model which can account for horizontal inhomogeneities 
in the plane of the ray path (described in \markcite{{\it Eyre,}} [1994]) and 
comparing observation vectors with the version of the forward model which 
assumes a horizontal homogeneous atmosphere [\markcite{{\it Palmer,}} [1998]]. 
Mid-latitude two-dimensional NWP fields (0$-$360$^{\circ}$) were used to 
simulate typical horizontal gradients in temperature and humidity. By 
considering small sections of the field at a time (typical of the horizontal 
resolution of radio occultation measurements which is of the order of 300~km), 
the complete field was traversed. Computing the ensemble mean from the 
difference between the two versions of the forward models allowed a reasonable 
estimate of the forward modelling error incurred by the spherical symmetry 
assumption to be computed. It is noted that this error estimate does
not represent the true error in observation space since the forward model
does not simulate the full error characterisation. Both 
\markcite{{\it Ahmad and Tyler,}} [1999] and \markcite{{\it Healy,}} [1999]
consider bending angle errors from horizontal gradients for specific cases.
However there is no published material that quantifies this error 
statistically. Simulation with a full 3-d raytracer through the UKMO mesoscale 
model fields suggest that the errors are approximately 3\% for ray paths near 
the surface, which is consistent with the value used in this work.

An error arises from representing an intrinsically high resolution problem 
with a crude resolution. This type of error is often called a 
representativeness 
error and describes the error from the inability of NWP model vertical grids 
to represent small-scale atmospheric structure, which are evident in GPS/MET 
RO measurements [\markcite{{\it Kursinski et al,}} [1997]]. The method used 
to estimate this quantity is described by \markcite{{\it Healy,}} [1998], and 
is found to be of the order of 2\% of the bending angle measurement in the 
troposphere and upper stratosphere, decreasing slightly in the middle 
stratosphere. This variation in the error is associated with the temperature 
variations in these region. 

The forward model used is based on geometric optics therefore does not account
for atmospheric diffraction. However, \markcite{{\it Kursinski et al,}} [1997] 
have shown that the geometric optics assumption is successful in describing 
propagation characteristics above a certain diffraction limit, and diffraction
is therefore not considered here.

\subsubsection*{Forward model parameter errors}

Uncertainties associated with the physical constants used to model the physical
system also cause modelling errors.
The major physical constants used in the forward model are the refractivity 
coefficients ($c_{1}$ and $c_{2}$ in equation \ref{eq:refractivity}) and the 
local radius of curvature.

The uncertainty of the refractivity coefficients do not represent the
error associated with their values but the uncertainty of the measured 
quantity; for this reason the information will not be included in the total 
observation error budget since it will result in a bias in the 
retrieval\footnote{Optimal estimation theory assumes that all the errors are 
unbiased.}.
The local radius of curvature assumed for zero altitude is a parameter that is 
used to compute the bending angle observations from atmospheric phase delay, 
therefore any error associated with this parameter will be present in the 
bending angle observations. This parameter is also used to compute geometric 
height levels from impact parameter levels. The uncertainty of this value is 
estimated to be approximately 100~metres 
[\markcite{{\it Kursinski et al,}} [1997]].

\subsubsection*{Total observation error}

The total observation error covariance ${\bf E}$ is constructed by 
adding the covariance matrices from observation noise, forward 
modelling and forward model parameters. 

The diagonal terms have also been constrained not to fall below a minimum 
value to account for the noisy upper stratosphere measurements.

Figure \ref{fig:bendingnoise} shows how the standard deviation values of the 
principal diagonal from each error contribute to the observation error 
covariance matrix.

The dominant source of error for the majority of the vertical range considered
is the forward modelling error, i.e.~representativeness error and horizontal 
inhomogeneity error, with the upper stratospheric noise limit providing the 
second largest contribution to the total 
error. In the upper stratosphere the total bending angle reverts to the upper
level minimum noise used. At near-surface altitudes, the forward model 
parameter error, i.e. local radius of curvature, is significant. The overall
effect from the local radius of curvature decreases exponentially due to the
hydrostatic relation.

\subsection{Convergence and quality control}

The method used to judge convergence relies upon values of the cost 
function, 
i.e.~if the relative change is smaller than a specified value (0.5\%) then 
the solution is determined to have converged. This method alone is found to
be a good indicator of convergence in this case.

For the work presented in this paper, the maximum number of iterations 
considered is 10; if the solution has not converged (determined by the 
method presented above within 10 iterations) then the calculation is halted 
and a numerical `flag' set. 

If, after convergence has been determined, the ${\bf J}({\bf x})$ 
value is greater than the $\chi^{2}$ value given the number of degrees of 
freedom at a set confidence level (in this case 99.9\%) then a numerical 
`flag' is set. Retrievals with flags set are omitted from any statistics.

Furthermore, to ensure the solution computed at each iteration is physical, 
the ln(specific humidity) elements are checked for super-saturation
and corrected if necessary.

\section{Results}
\label{sect:results}

In this section the performance of the optimal estimation retrieval scheme 
is examined using simulated profiles and realistic error estimates. 
For each simulated profile, a `true' profile is established by taking
one of a set of profiles of UKMO model analyses from which to compute the
`true' observation vector. The associated background atmospheric profile is 
calculated by perturbing the `true' profile thus

\begin{equation}
{\bf x}^{b} = {\bf x}^{t} + \sum_{i=1}^{n} \epsilon_{i} \lambda_{i}^{1/2} 
              {\bf P}_{i}
\label{eq:simx}
\end{equation}
where the superscript $t$ denotes the `truth', $\lambda_{i}$ and 
${\bf P}_{i}$ are the $i$th eigenvalues and eigenvectors of the forecast
error covariance matrix and $\epsilon_{i}$ represents the $i$th number 
drawn from a normal distribution of random numbers.

Observation noise is modelled and superimposed onto the observation vector 
using the method analogous to equation \ref{eq:simx}, utilising the 
eigenvector and eigenvalues from the total observation error covariance matrix.

\subsection{Ensemble of numerical simulations}

Using the method described by equation \ref{eq:simx}, simulated observations
and realistic background profiles were produced. These were used as inputs to 
the inversion scheme to obtain retrieved profiles which were subsequently
compared with the `true' profiles to assess the impact of the observations on 
the background information.

The observation level values (i.e. impact parameter, local radius of
curvature and geographical position) 
have been taken from data during `prime-times'\footnote{Periods of time when 
the received signals are free from anti-spoofing military encryption.} 
1 and 2 [\markcite{{\it Rocken et al,} 1997}], in an effort to simulate 
realistic retrieval scenarios. The latitudinal and longitudinal distribution 
of these occultation events (determined by GPS sampling) are varied, thus 
providing a mixed ensemble of polar, tropical and mid-latitudinal 
occultation events.

Five hundred profiles with random temperature, humidity and surface pressure 
conditions have been tested, and successful retrievals (i.e.~which pass 
quality control) were obtained in all 
but eight cases. In most cases convergence is obtained within three or 
four iterations. In general, the ${\bf J}({\bf x})$ values at convergence were 
comparable to the number of degrees of freedom considered, as expected 
[\markcite{{\it Marks and Rodgers,} 1993}]. This suggests that although the
$\chi^{2}$ quantity is only strictly valid for linear problems it can be used 
reliably as a quality control for the retrieval.

The eight cases which do not pass the specified quality control have been 
examined. They are found to be cases in which the cost function has been 
minimised successfully but the converged value is too large compared with the 
$\chi^{2}$ distribution. These spurious converged profiles represent 
artificial
outliers, which are generated when the increment described by equation 
\ref{eq:simx} is large enough to make the inverse problem grossly non-linear. 
A small number of profiles are expected to have this problem due to the normal 
distribution of random numbers used in the method of simulating atmospheric 
profiles. The profiles which failed the quality control have not been included 
in the statistics shown.

For each successful retrieval, the retrieval error and the background error
(first-guess error) have been calculated, and the mean and standard deviation
values of these data have been computed. The standard deviation values 
represent the errors ascribed to each element of the solution and background 
state vector, and can be compared directly with the square root values of the 
principal diagonal of the background error covariance matrix assumed in the 
retrieval.
The ratio of the retrieval error estimates to the forecast error estimates is 
related to the amount of information the measurements supply to the NWP 
system. 

An improvement vector is defined which indicates how the retrieval has 
improved the knowledge of the background state throughout the atmospheric 
profile, and will be used to complement the r.m.s.~statistics presented. The 
improvement vector ${\mbox{\boldmath{$\eta$}}}$ is given by

\begin{equation}
{\mbox{\boldmath{$\eta$}}}_{j} =  100 \times \left[ 1 - 
           \left(\frac{{\bf \hat{S}}_{j,j}}{{\bf C}_{j,j}}\right)
           ^{\frac{1}{2}}\right]
\label{eq:imprv-vectr}
\end{equation}
where $j$ is the matrix and improvement vector element index, and all other 
variables are as before.

\begin{figure}[tbh!]
\centerline{
\epsfysize=12cm
\epsffile{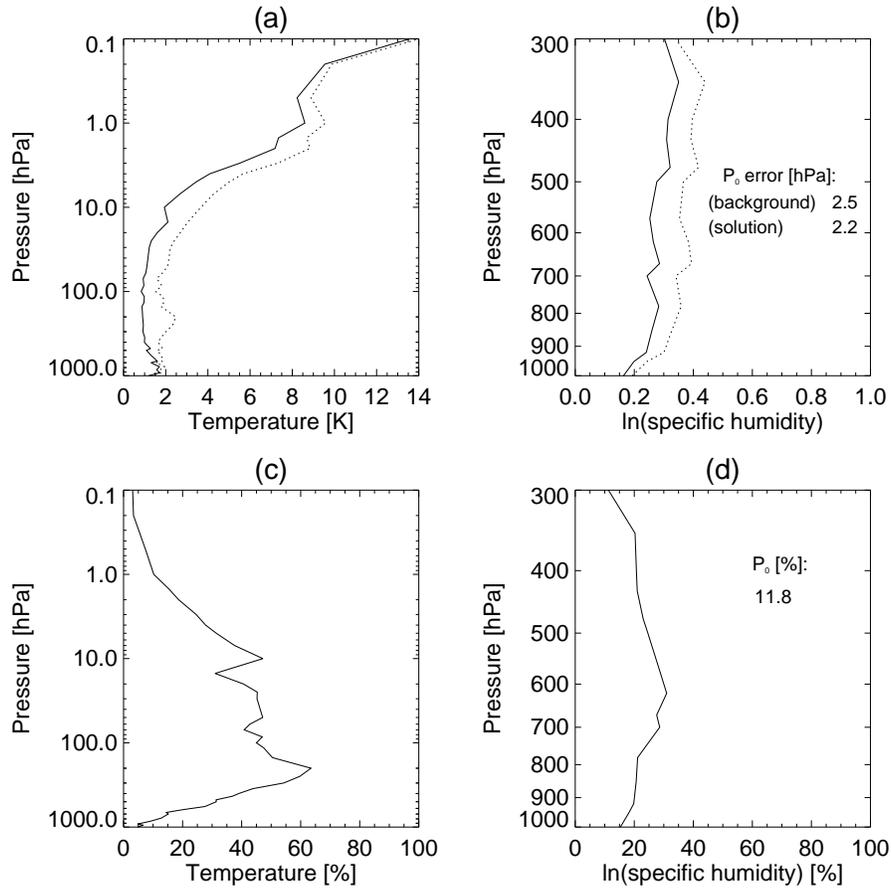}}
 \caption{Theoretical error estimates from the ensemble of simulated 
          retrievals. The upper panels
          show the theoretical r.m.s~errors, where the solid and dotted lines 
          represent the retrieval and background errors, respectively. Panel 
          (a) shows the errors from the temperature solution and the panel (b)
          shows the errors from the $\ln$(specific humidity) solution with 
          the surface pressure solution error inset. The lower panel show 
          the corresponding improvement vector. Panels (c) and (d) show the 
          temperature and $\ln$(specific humidity) elements with the 
          surface pressure improvement inset of panel (d).}
\label{fig:simsx}
\end{figure}

The results from the ensemble of simulated retrievals are summarised by figure
\ref{fig:simsx}.
The upper panels ((a) and (b)) show the computed r.m.s.~errors from the the
simulated retrievals. The forecast errors resemble those shown by figure 
\ref{fig:ukmoerror} as expected, modified slightly by the modelled 
observation noise.

The temperature improvement vector suggests that optimal estimation 
considerably improves upon the prior knowledge of the atmospheric temperature, 
from the lower-troposphere to the mid-stratosphere. The temperature 
improvement vector declines in the upper stratosphere partly because of the 
upper noise limit used, which is comparable to forecast errors in 
observation space, and partly because it represents the lower limit of the 
observation 
vectors used in the ensemble of simulations. The temperature improvement 
vector declines in the lower-troposphere where humidity becomes more 
significant, and both quantities are retrieved simultaneously, thus the 
emphasis is shifted from temperature to $\ln$(specific humidity). It is clear 
from the plot that there is a gradual decline in the temperature retrieval 
quality from 300 to 1000~hPa as the humidity retrieval quality improves, where
the background uncertainty information is being used to resolve the 
temperature-specific humidity ambiguity in the refractivity. 

Both the temperature and humidity knowledge decline near the surface because 
the majority of occultation events presented here terminate typically above
one kilometre.

The results shown by figure \ref{fig:simsx} confirm that the method is 
suitable for the purpose of non-linear optimal estimation using RO 
measurements. Together, the theoretical r.m.s.~errors and the computed
improvement vectors suggest that there are improvements in the prior 
knowledge of the atmospheric state from near-surface to the upper stratosphere.
In particular, the level of surface pressure improvement suggests that the 
RO observations can improve the prior knowledge of 
the surface pressure.

\subsection{Solution error characterisation}

\begin{figure}[tbh!]
\centerline{
\epsfysize=8cm
\epsffile{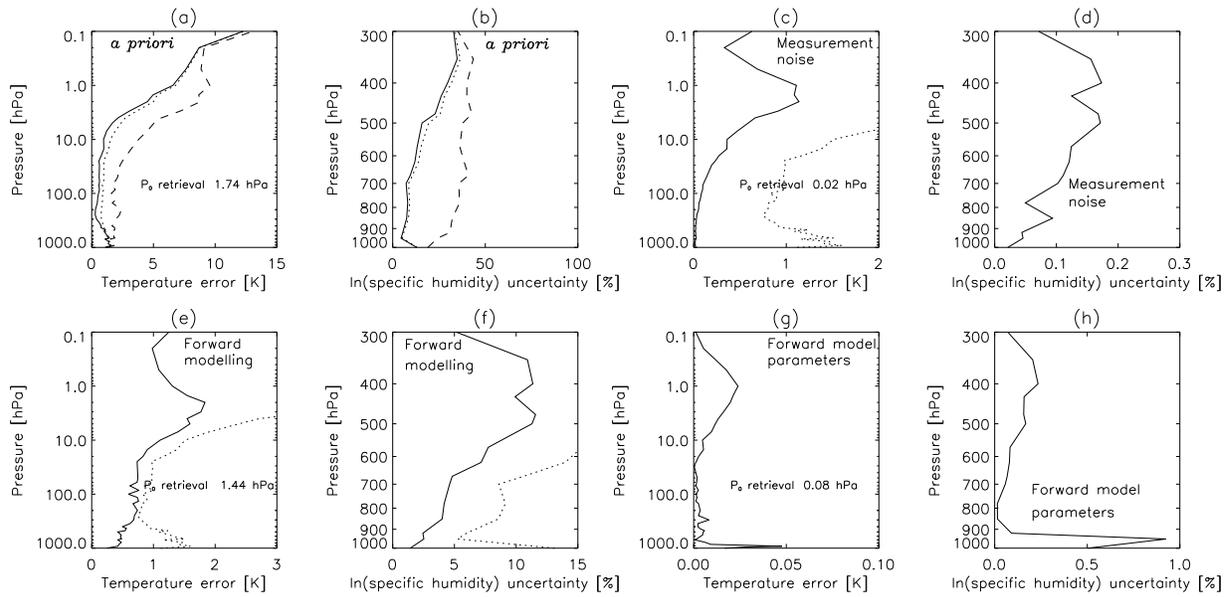}}
 \caption{Solution error characterisation of the optimal estimation inverse 
          method for the example occultation profile. Left panels of a pair 
          show the contribution to the temperature retrieval 
          error and the right panels show the error contributions to the 
          $\ln$(specific humidity) retrieval error. The surface pressure 
          retrieval error is inset of the left panels. Panels (a) and (b) 
          show the {\it a~priori} contribution to the retrieval error, where
          the dashed shown represent the standard deviation values of the 
          principal 
          diagonal of the {\it a~priori} error covariance matrix;  
          (c) and (d) the measurement noise contribution; (e) and (f) the 
          forward modelling contribution; and (g) and (h) the 
          forward model parameter contribution. The total solution error is
          superimposed on all panels (dotted line) to provide an indication as
          to which contributions are prominent.}
\label{fig:ctbn-optimal}
\end{figure}

Using the optimal estimation inverse theory outlined in section 
\ref{sect:optimal} an error analysis can be obtained which allows a full 
characterisation of the solution (for a detailed account see 
\markcite{{\it Rodgers,}} [1990]). The error associated with the solution 
vector can be split into its constituent parts, namely error from the 
background error estimates, forward modelling, forward model parameters and 
observation noise. A mid-latitude retrieval\footnote{Henceforth will be 
referred to as the example occultation profile.} (which spans 0.7$-$60~km) 
has been used as an example to illustrate the method 
(figure \ref{fig:ctbn-optimal}). 

Panels (a) and (b) show that the {\it a~priori} provides almost all the
information to the temperature and $\ln$(specific humidity) retrieval 
above 10~hPa and 500~hPa, respectively. 

Panels (c) and (d) show the observation noise contribution to the retrieval 
error which is comparatively small. 

Panels (e) and (f) show the forward modelling contribution to the 
retrieval error. The structure of this contribution is very similar to that of
the observation noise, but has a larger associated error. In general, forward 
modelling represents the second largest contribution to retrieval error.

The temperature solution error contributions shown by panels (c) and (e) peak 
at 3~hPa, at the point where the stratospheric noise limit contribution to 
the total observation error budget peaks (figure \ref{fig:bendingnoise}).
The observation contribution to the solution error begins to increase above
about 0.3~hPa. Above this pressure level the {\it a~priori} increases more 
rapidly, and so the observation is given more weight, resulting in a larger
contribution to the solution error.

Panels (g) and (h) show that forward model parameter error represents the
smallest contribution to the total retrieval error. This contribution to the
temperature and  $\ln$(specific humidity) solutions peak near 1000~hPa due to 
the local radius of curvature error. The contribution to the surface pressure 
solution represents a small fraction of the total retrieval error.

Figure \ref{fig:ctbn-optimal} indicates that the dominant contributions to the
solution error are from the {\it a~priori} and forward modelling, 
suggesting that particular efforts should be made to improve their accuracies.
It should be noted that the solution error depends on the assumptions made
about the uncertainty statistics.

\subsection{Quality of surface pressure retrievals}

\begin{figure}[tbh!]
\centerline{
\epsfysize=12cm
\epsffile{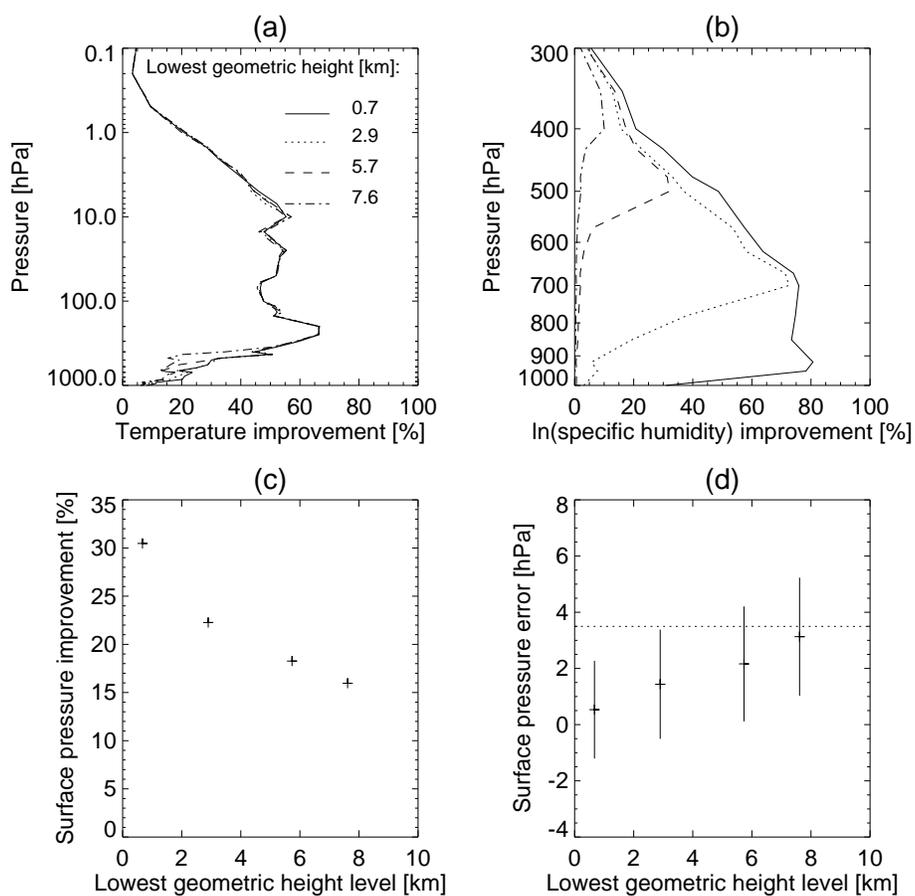}}
 \caption{Investigating the relationship between the quality of surface 
          pressure retrievals and the lowest geometric height level of the
          occultation using the example occultation profile. Panel (a), (b) 
          and (c) show the temperature, 
          $\ln$(specific humidity) and surface pressure improvement 
          vectors; and panel (d) shows the difference (with errors bars) 
          between the truth and the retrieval. The broken line in panel (d)
          represents the difference between the `true' and the assumed
          value for the surface pressure.}
\label{fig:p0simulations}
\end{figure}

It is found that the quality of the surface pressure retrievals is dependent
on the vertical extent of the occultation, i.e.~how closely it approaches the 
surface. To illustrate this point, the example occultation profile is used. By 
systematically removing observations near the surface and re-retrieving 
temperature, $\ln$(specific humidity) and surface pressure, it can be shown 
how the vertical range of the occultation is important to the quality of the 
surface pressure retrieval (figure \ref{fig:p0simulations}). 

Panels (a), (b) and (c) show the temperature, $\ln$(specific humidity) and
surface pressure improvement vectors get smaller in the lower atmosphere 
with increasing values for the lowest geometric height level, as expected. 
Panel (d) shows that the retrieval which includes all the observations 
provides enough information to retrieve accurately the true value for surface
pressure; as the number of near-surface observations decreases the retrieval 
becomes smaller and smaller, approaching the prior value. 

It is interesting to note that the surface pressure information does not 
decrease as quickly as expected. Indeed there is still a considerable 
amount of surface pressure information towards the upper troposphere. 
This variation in surface pressure information is due to the
link between height and pressure through the hydrostatic relation.

It can be concluded from this experiment that using optimal estimation, radio 
occultation measurements of the Earth possess surface pressure
information even if the occultation has been completed in the 
mid-troposphere (e.g.~due to atmospheric multipath interrupting the transmitted
signal). 
This is important to appreciate when validating surface pressure retrievals 
using real data.

\subsection{Sensitivity to inverse model statistics}

In this section we present the results from a sensitivity study in which
the statistics used to compute the optimal estimate are changed in order to 
investigate the retrieval sensitivity to such alterations. The three
statistics that are altered are the observation noise, the forward modelling
error and the forecast error since these represent  the largest contributions
to the solution error as shown by figure \ref{fig:ctbn-optimal}.

Altering the observation noise may simulate the possible changes in 
observation noise sources, e.g.~improved high-order ionospheric or poor 
quality clocks aboard the GPS receivers. 

Changing the forward modelling error is a crude method of simulating 
the possibility of assimilating intrinsically high resolution GPS radio 
occultation measurements with lower or higher resolution NWP model fields,
and/or an occultation through a frontal system or a relatively horizontally
inhomogeneous atmosphere. 

Altering the prior error covariance matrix represents the effect of changing 
the prior knowledge of the atmosphere. Increasing these error estimates can 
represent the extent of the knowledge of the atmosphere in the southern 
hemisphere where other atmospheric information is sparse. Decreasing this error
may represent a more realistic model dynamics/climatology and/or greater
confidence in other similar atmospheric observations used to initialise the 
model. For this particular test the variance values are changed, whilst 
retaining the existing correlations.

\begin{figure}[tbh!]
\centerline{
\epsfysize=14cm
\epsffile{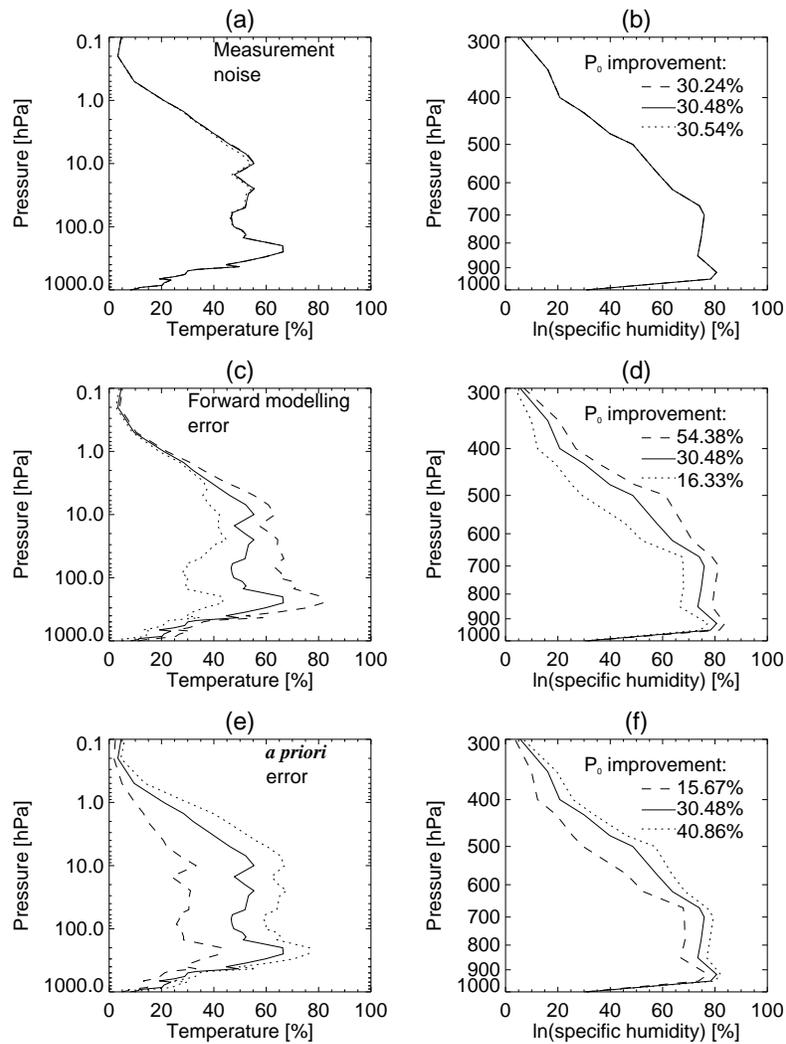}}
 \caption{Improvement vectors from the retrieval sensitivity studies using the
          example occultation profile. In 
          general, left panels show the temperature improvement vector 
          elements and the right panels show the $\ln$(specific humidity)
          improvement vector elements, with the surface pressure improvement
          vector element inset. For panels (a)$-$(f), dashed curves represent 
          a reduction in the quantity altered, dotted curves represent an 
          increase in the quantity altered and solid curves represent the 
          control case.
          Panels (a) and (b) show the results from doubling/halving the 
          measurement noise; panels (c) and (d) show the results from 
          increasing and decreasing the representative error by $\pm$50\%;
          and panels (e) and (f) show the results from increasing and 
          decreasing the principal diagonal standard deviation errors from the 
          {\it a~priori} error covariance matrix by $\pm$50\%.}
\label{fig:simulations}
\end{figure}

Figure \ref{fig:simulations} shows the improvement vectors from the retrieval
sensitivities outlined above. 
Improvement vectors are used to present the results of this study because it
is the relative improvement on the prior estimate of the atmospheric state
that we are interested in.

Panels (a) and (b) shows that doubling or halving the observation noise
has little effect on the overall improvement, reflecting the contribution from
this error source to the total observation error covariance matrix (figure 
\ref{fig:bendingnoise}). The changes due to $\ln$(specific humidity) are very 
slight. The surface 
pressure improvement changes typically by a few percent.

Panels (c) and (d) correspond to increasing or decreasing the forward modelling
error (by 50\%). This error source provides the largest contribution to the 
total observation error, and as such has a large influence on the degree of 
improvement. The difference in the temperature improvement is of the order of 
15\% in the upper troposphere and lower stratosphere, above which other error 
sources are more important, and below which (the lower troposphere) the 
emphasis is shifted from the temperature retrieval to the 
$\ln$(specific humidity) and surface pressure retrieval. The improvement 
response for $\ln$(specific humidity) is less pronounced than for temperature 
peaking at $\pm$8\%. The surface pressure improvement variation is 
approximately $\pm$15\%.

Panels (e) and (f) shows that increasing or decreasing the standard deviation
of the prior error increases or decreases the improvement of temperature, 
$\ln$(specific humidity) throughout the vertical range of the observations as 
expected. 
Decreasing the {\it a~priori} error means better background knowledge, 
consequently the weighting of the {\it a~priori}/observation is 
increased/decreased. This corresponds to a small improvement relative to the
background atmospheric knowledge. Positive and 
negative temperature improvement differences are of the order of 20\% 
throughout the range described by the observation vector; the 
$\ln$(specific humidity) improvement is of the order of $\pm$10\%; and the 
surface pressure improvement is of the order of $\pm$10\%.

This sensitivity study has looked at some of the extreme case scenarios in 
which the statistics assumed for the optimal estimation inverse method
have been changed. It has been shown that the retrieval method is most
sensitive to the background errors and forward modelling errors; the latter
being related to errors due to the spherical symmetry assumption. 
The results from increasing the background error estimates are especially
important to note since they represent a real possibility when dealing with 
any reasonable measurements in the data sparse southern hemisphere.

\section{Conclusions}
\label{sect:disc}

In this paper we have demonstrated a prototype optimal 
estimation inverse method for GPS radio occultation observations. 

The method is a non-linear, maximum {\it a posteriori} technique which can 
accommodate most aspects of the real radio occultation problem. 
In particular, it is able to account for some of the error
incurred from assuming local spherical symmetry which is not possible using the
analytic inverse method. The optimal estimation technique handles the 
temperature$-$water vapour ambiguity in a more rigorous way, rather than the 
sub$-$optimal manner inherent with the analytic inverse method.

The optimal estimation inverse method is used here as an iterative method but 
is found to be stable and to converge rapidly in most cases. 
The value of the cost function at each iteration can be used reliably to 
judge convergence and as an indicator of sensible results, allowing 
anomalous retrieval conditions to be recognised and omitted, thus preventing 
gross retrieval errors. 

The method is shown to be suitable for retrieving values for surface pressure. 
Hence, this method of utilising radio occultation observations of the Earth's
atmosphere has the potential to improve both atmospheric and oceanographic 
models, which may lead to improved predictions of the weather and climate.
The quality of the surface pressure retrieval is shown to depend on the
vertical extent of the occultation, i.e.~higher quality retrievals are 
attainable with occultations that reach low altitudes. 

It should be noted that the background statistics assumed in the paper 
represent global statistics. However, for purposes of demonstrating a 
prototype retrieval scheme for radio occultation observations they are found
to be adequate. It has been shown that the retrieval accuracies, and hence 
the weight which should be given to the data in the subsequent model analysis, 
are sensitive to both forecast error uncertainties and values used to describe 
the forward modelling error.

%
%

\acknowledgments
The authors would like to thank the Jet Propulsion Laboratory for 
providing the GPS/MET radio occultation dataset, J. P. Luntama for
the bending angle error estimates, and C. D. Rodgers, A. Dudhia and  H.
Roscoe for comments on earlier drafts.

%
%

\end{article}

\newpage

%
%

\end{document}